# WU-NEAT: A clinically validated, open-source MATLAB toolbox for limited-channel neonatal EEG analysis

Zachary A. Vesoulis, Paul G. Gamble, Siddarth Jain,
Nathalie M. El Ters, Steve M. Liao, Amit M. Mathur

*Abstract— Goal*: Limited-channel EEG research in neonates is hindered by lack of open, accessible analytic tools. To overcome this limitation, we have created the Washington University-Neonatal EEG Analysis Toolbox (WU-NEAT), containing two of the most commonly used tools, provided in an open-source, clinically-validated package running within MATLAB. *Methods*: The first algorithm is the amplitude-integrated EEG (aEEG), which is generated by filtering, rectifying and time-compressing the original EEG recording, with subsequent semi-logarithmic display. The second algorithm is the spectral edge frequency (SEF), calculated as the critical frequency below which a user-defined proportion of the EEG spectral power is located. The aEEG algorithm was validated by three experienced reviewers. Reviewers evaluated aEEG recordings of fourteen preterm/term infants, displayed twice in random order, once using a reference algorithm and again using the WU-NEAT aEEG algorithm. Using standard methodology, reviewers assigned a background pattern classification. Inter/intra-rater reliability was assessed. For the SEF, calculations were made using the same fourteen recordings, first with the reference and then with the WU-NEAT algorithm. Results were compared using Pearson's correlation coefficient. *Results*: For the aEEG algorithm, intra- and inter-rater reliability was 100% and 98%, respectively. For the SEF, the mean±SD Pearson correlation coefficient between algorithms was 0.96±0.04. *Conclusion*: We have demonstrated a clinically-validated toolbox for generating the aEEG as well as calculating the SEF from EEG data. Open-source access will enable widespread use of common analytic algorithms which are device-independent and not subject to obsolescence, thereby facilitating future collaborative research in neonatal EEG.

## I. Introduction

MAYNARD and Prior first developed the analog limited-channel cerebral function monitor (CFM) in the 1960s [1]. The original analog technology was subsequently updated to a digital implementation and validated in neonates by de Vries, Hellström-Westas and others in the 1980s-90s [2]–[4]. Current devices are compact, easy to use, and the electrodes have an excellent safety profile for even the most fragile of preterm skin [5]. The recording can then be analyzed both qualitatively and quantitatively utilizing proprietary software sometimes provided by the manufacturer. As a result, this technology has been used for the study of seizures in premature infants [6], maturation changes in EEG background classification over the course of hospitalization [7], and the early detection of seizures in term infants with hypoxic-ischemic encephalopathy (HIE) [8].

Qualitative analysis is performed by visual interpretation of the amplitude-integrated EEG, a display derived from the raw EEG signal, allowing the clinician to evaluate cerebral health on a continuum, from normal voltage with sleep-wake cycling, to burst-suppression, to flat tracing. Several standardized methods for interpretation have been developed [9], [10]. The digital nature of the recordings also lends itself well to quantitative analysis, ranging from simple measures such as minimum, maximum and mean values, to complex frequency domain techniques evaluating spectral power [11], [12].

Limited-channel EEG monitors with specific application to the neonate were developed by several manufacturers (BrainZ, Olympic, Nicolet) before eventual market consolidation. This has left a field littered with a number of different proprietary devices, each of which has a unique and incompatible file type. Further compounding this problem is a limited interest in further developing or updating the analysis software by manufacturers (as there is a limited commercial market), necessitating the continued use of software a decade, or more, old. A final challenge is the lack of established specifications for quantitative algorithms, beyond generalities. Heterogeneities in algorithm function may lead to considerable variance in the output, leading to considerable challenges when comparing results collected using devices from different manufacturers.

We have addressed these problems by developing WU-NEAT (Washington University Neonatal EEG Analysis

This manuscript was submitted on March 22, 2018. This work was supported in part by the 1. The Washington University Institute of Clinical and Translational Sciences KL2 Training Program (NIH/NCATS KL2 TR000450), The Barnes-Jewish Hospital Foundation and the Washington University Institute of Clinical and Translational Sciences Clinical and Translational Funding Program (NIH/NCATS UL1 TR000448), and the Thrasher Research Fund.

*Z.A. Vesoulis, P.G. Gamble, N.M El Ters, S.M. Liao, and A. M. Mathur are with the Department of Pediatrics, Division of Newborn Medicine. Washington University School of Medicine, St. Louis, MO, USA. S. Jain is with the Department of Neurology, Division of Pediatric Neurology, Washington University School of Medicine, St. Louis, MO, USA. (correspondence e-mail: vesoulis_z@wustl.edu).

Toolbox), an open source EEG toolbox, designed for the analysis of limited-channel EEG recordings of neonates. This set of tools addresses the shortcomings of the current research environment by recreating two key functions of the original aEEG software and providing the framework for development of additional tools. WU-NEAT runs within the widely available MATLAB scientific programming language (The Math Works, Natick MA). As open-source software, the source code for the software is freely available for non-commercial use, reducing barriers to performing quantitative EEG analysis, allowing inspection of the inner-workings of the algorithms, and standardizing the framework for quantitative EEG, facilitating inter-center comparison.

The key contributions of WU-NEAT are that it:
1. Defines a standardized, open file format for the storage of limited-channel EEG files, enabling long-term interoperability and to prevent obsolescence.
2. Establishes a core framework which enables access to stored files, a modular plugin interface for addition of analytic tools, interactive display of analytic output, and export of results.
3. Implements the two most frequently used tools for analysis of the limited channel EEG in the neonate – amplitude integrate EEG (aEEG) and spectral edge frequency (SEF).

*WU-NEAT can be downloaded from:*
*http://research.peds.wustl.edu/neat.*

## II. MATERIALS AND METHODS

### A. EEG recording

EEG recordings used in the development and testing of WU-NEAT were prospectively recorded from preterm infants born before 28 weeks gestation and term infants with hypoxic-ischemic encephalopathy (HIE). A common lead placement technique was used for all infants (C3-P3, C4-P4) and all recordings were made using the BrainZ BRM3 aEEG monitor (Natus Medical, San Carlos, CA). The EEG signals are sampled at 64 Hz, while the impedance is sampled at 1 Hz. The recordings of the preterm infants were 72 hours in length (starting at birth) and the recordings of HIE infants were 96 hours in length (encompassing the entirety of therapeutic hypothermia treatment and rewarming).

The BRM3 generates recordings in the *.BRM binary file format, which cannot be opened by any software package other than the one originally provided by the manufacturer (AnalyZe, Natus Medical, San Carlos, CA). AnalyZe does permit export of the raw two-channel EEG, two-channel impedance recording, and the calculated two-channel aEEG traces as an Excel (Microsoft, Redmond, WA) spreadsheet, although owing to the 32-bit nature of the software, it is limited to export of 65536 rows at a time (approximately 17 minutes of data at 64 Hz). For the validation component, EEG recordings from 14 infants were fully exported using this method.

### B. File-format specification

As noted, a significant weakness of the current research environment is the lack of a standardized, open file format. Incorporated in this project, we propose a new file format specification which will enable permanent accessibility and guaranteed interoperability between institutions and across computing platforms. For the EEG signals, the comma separated value (CSV) file format will be used to generate a four-column matrix, with the date in column 1 (YYYY-MM-DD), time in column 2 (HH:MM:SS.FFF), the left raw EEG signal (in µV) in column 3, and the right raw EEG recording (in µV) in column 4. A separate CSV file with four columns will be used to store impedance information with date and time stored in columns 1 and 2, left impedance (kΩ) in column 3, and right impedance (kΩ) in column 4. WU-NEAT reads properly formatted CSV files and imports them into MATLAB as datetime or numeric arrays with appropriate labeling.

### C. Core interface

The tools contained within WU-NEAT can be used as freestanding functions within MATLAB or can be used within a graphical user interface (GUI) which combines data import functions, display of the EEG signal, and selection of data regions for analysis. An example of the interface is shown in Figure 1. The interface is intentionally flexible and modular, allowing for later addition of other analytic tools.

The EEG display functionality contains two components, enabling navigation of the recording and synchronized access of the corresponding raw trace to a selected time point in the aEEG signal. Simultaneous use of raw and aEEG signals is important to improve the specificity of seizure diagnosis and to rule out artifact as the source of a sudden increase in the aEEG baseline [13].

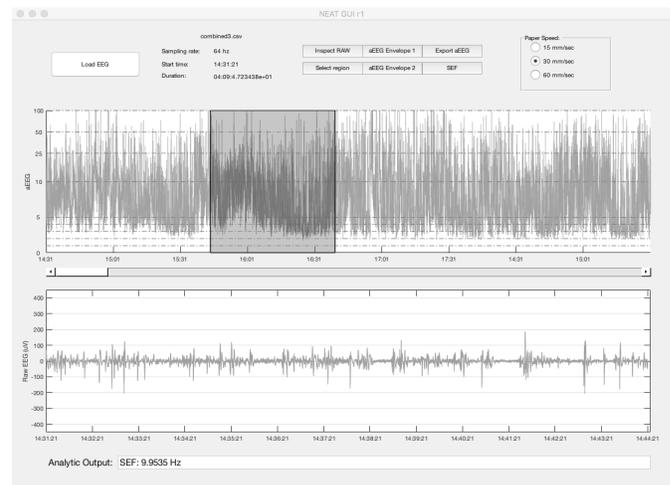

Fig. 1. Screenshot of the graphical user interface with linked display of the raw trace and associated aEEG and the calculated SEF of the selected region.

### D. aEEG algorithm

The amplitude-integrated EEG fundamentally is a transformation of the raw EEG signal, one which accentuates EEG activity within the frequency band of greatest interest in the neonate (2-15 Hz) and is time compressed, enabling rapid, longitudinal evaluation of background activity as well as the presence or absence of seizures. The underlying approach to calculation, and interpretation, of the amplitude-integrated EEG has been well described in both analog [1] and digital [14] implementations and has been reproduced in WU-NEAT.

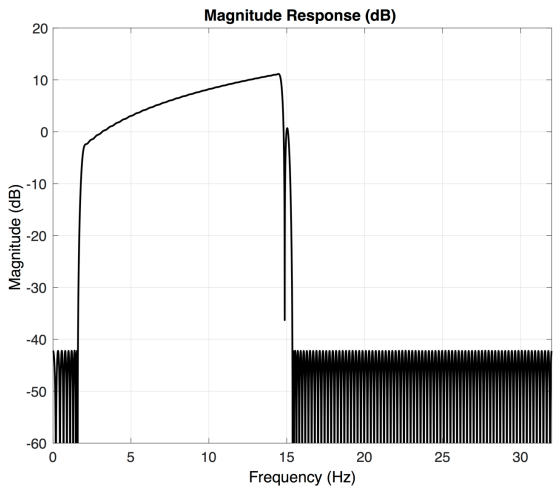

Fig. 2. Frequency response of the asymmetric bandpass filter used in the aEEG algorithm, which filters out the signal below 2 Hz and above 15 Hz..

To create the aEEG signal, the raw EEG is first passed through an asymmetric pass-band filter ($300^{th}$-order Parks-McClellan linear-phase FIR filter) which strongly attenuates the signal below 2 Hz (low-frequency physiological artifact) and above 15 Hz (muscle activity, line noise, respiratory and heart rate artifact). The pass-band (2-15 Hz) undergoes gradual amplification with a slope of 12 dB/decade, to compensate for the diminished amplitude of higher frequency EEG activity [14], [15]. The frequency response is shown in Figure 2.

After pass-band filtering, the signal is rectified and the signal envelope is generated by use of a low-pass Butterworth filter using zero-phase filtering (Figure 3). The resulting transformed signal is displayed in a time-compressed fashion, generally with 3.5hours of data displayed at a time. In order to accentuate the baseline, a key feature necessary for determination of the aEEG background, the y-axis is a linear-log scale—linear scaling from 0-10 μV and logarithmic from 10-100 μV.

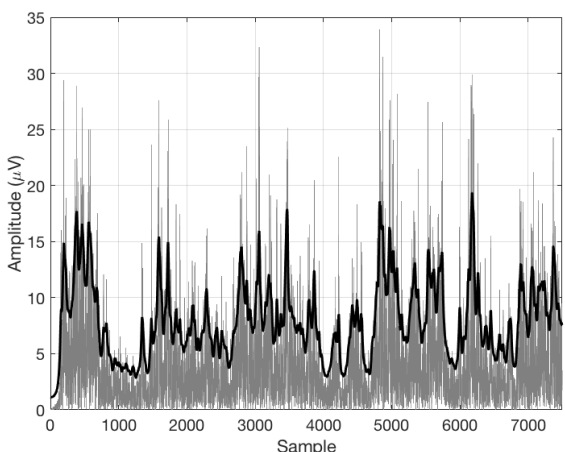

Fig. 3. The output of the enveloping function (the aEEG signal) is shown in black, on top of the filtered, rectified raw EEG signal (gray).

### E. Spectral edge frequency

Power spectral analysis, a method for quantifying the strength of oscillations in a given frequency band has been a commonly used method for signal processing since the development of a digital algorithm by Welch [16] in 1967. By quantifying the distribution of spectral power in an EEG recording, Bell et al. were able to delineate [11] the frequency below which 95% of the power resides, a method termed spectral edge frequency (SEF). This approach has been used to confirm the qualitative observation that mean frequency of EEG activity increases with increasing gestational age [7] and can identify delayed maturation in infants with brain injury [12], [17].

A flexible tool for calculating the SEF has been incorporated into WU-NEAT. First, the algorithm partitions the input signal into non-overlapping segments of 60 seconds (3840 samples) in length. In order to eliminate the effects of low-frequency physiologic and high-frequency respiratory, cardiac, and line noise artifact, the selected region of raw EEG signal undergoes pass-band filtering between 2 and 20 Hz using a $5^{th}$-order Butterworth filter. The total spectral power of the filtered EEG signal is then calculated using Welch's power spectral density estimate, which divides the signal into 8 segments with 50% overlap. A Hamming window function is applied to reduce spectral leakage. The $SEF_x$ is then calculated at the frequency below which $x$% (user defined value) of the spectral power is present.

### F. Validation

All studies procedures were reviewed and approved by the Human Research Protection Office at Washington University. All statistical testing was performed in R version 3.3.3 (The R Foundation for Statistical Computing, Vienna, Austria).

#### 1) Sample data

A sample dataset of preterm infants, born before 28 weeks who underwent 72 hours of continuous aEEG monitoring were utilized for validation in the preterm infant. A second dataset of term infants with hypoxic-ischemic encephalopathy who underwent 96 hours of continuous aEEG monitoring were utilized for validation in the term infant. All recordings were derived from previously conducted observational studies.

#### 2) aEEG validation

As the amplitude-integrated EEG is fundamentally a qualitative tool, validation proceeded qualitatively. For the preterm and term datasets, the aEEG of a 4-hour segment of raw EEG data was calculated first using the algorithm from the manufacturer-supplied software (AnalyZe) and then again using the algorithm in WU-NEAT. The resulting signals were plotted within MATLAB to prevent subtle differences in non-signal plot details from biasing interpretation. Three authors not involved in the creation of the algorithm (SML, NE, AMM), and who have significant experience in the interpretation of preterm and term aEEGs, were given the resulting plots in random order. They were asked to score the aEEG background using the 5-category (continuous with cyclicity, continuous, discontinuous, burst suppression, flat trace) system developed by Hellstrom-Westas et al. [10]. The percent agreement within raters (to compare output of WU-NEAY and AnalyZe) and between raters (to assess agreement on underlying background pattern) was calculated using Fleiss' Kappa. All statistical work was perfomed using R v 3.3.3 (R Foundation for Statistical Computing, Vienna, Austria).

TABLE I
SAMPLE CHARACTERISTICS

|  | Preterm cohort (n=6) | Term cohort (n=8) |
|---|---|---|
| Gestational age, mean (SD), weeks | 25.8 (0.9) | 38.2 (2.0) |
| Birth weight, mean (SD), grams | 845.0 (244.7) | 2968.0 (637.0) |
| Male sex, n (%) | 3 (50) | 3 (38) |

*3) SEF validation*

Using the same 4-hour segments of data as in the aEEG validation, the spectral edge frequency was calculated, once using the AnalyZe algorithm and again using the algorithm in WU-NEAT. As the recordings were 4 hours in length, a total of 240 one-minute epochs were compared for each infant. The degree of correlation within each subject was assessed using the Pearson correlation coefficient and averaged across the entire validation cohort.

### III. RESULTS

*A. Sample characteristics*

A total of 14 infants were selected for the validation cohort (preterm n=6, term n=8). Mean±SD EGA was 25.8±0.9 weeks for the preterm cohort and 38.2±2.0 weeks for the term cohort; mean±SD birthweight was 845.0±244.7 grams for the preterm cohort and 2968.0±637.2 grams for the term cohort (Table 1).

*B. aEEG validation*

aEEG traces were successfully produced for all 14 included infants. Examples of the aEEG output from AnalyZe and WU-NEAT are shown in Figure 4. Given that there were three raters and 14 recordings, a total of 42 scores were generated. For the 14 recordings, the predominate background pattern was continuous in 4/14 (28%), discontinuous in 5/14 (36%), and burst suppression in 5/14 (36%). There was complete intra-rater agreement; raters scored all 14 examples with the same background classification when viewing both AnalyZe and WU-NEAT generated images. There was also substantial agreement between raters, with disagreement on the background pattern in only one infant, κ=0.65. In this case, the aEEG tracing generated by WU-NEAT was scored as discontinuous by two raters and burst suppression by the third rater.

*C. SEF validation*

The SEF$_{95}$ was successfully generated for all 14 infants using the AnalyZe and WU-NEAT algorithms. There was strong correlation between the results, with a mean Pearson correlation coefficient of 0.96±0.04.

### IV. DISCUSSION

Although there is obvious research potential of quantitative, limited-channel EEG in the neonate, the currently available tools for analysis are either non-existent, woefully outdated, or do not directly address the needs of researchers. With WU-NEAT, we have overcome these challenges using an open-source approach, allowing for inspection of the underlying methodology, the ability to add additional tools over time, all in a free, platform-independent package.

Other tools for quantitative EEG analysis are available, but do not sufficiently address the current software gap for limited-channel neonatal EEG. The most prominent example is EEGLAB [18], a MATLAB-based software package which contains many quantitative tools. However, EEGLAB was designed for use with conventional (or high-density) EEG recordings of adults. The manner of these recordings is very different from limited-channel EEG in the neonate, and thus many of the built-in functions are inaccessible. Adapting WU-NEAT as an EEGLAB plugin would add additional software dependencies and potential compatibility issues without adding any significant benefit.

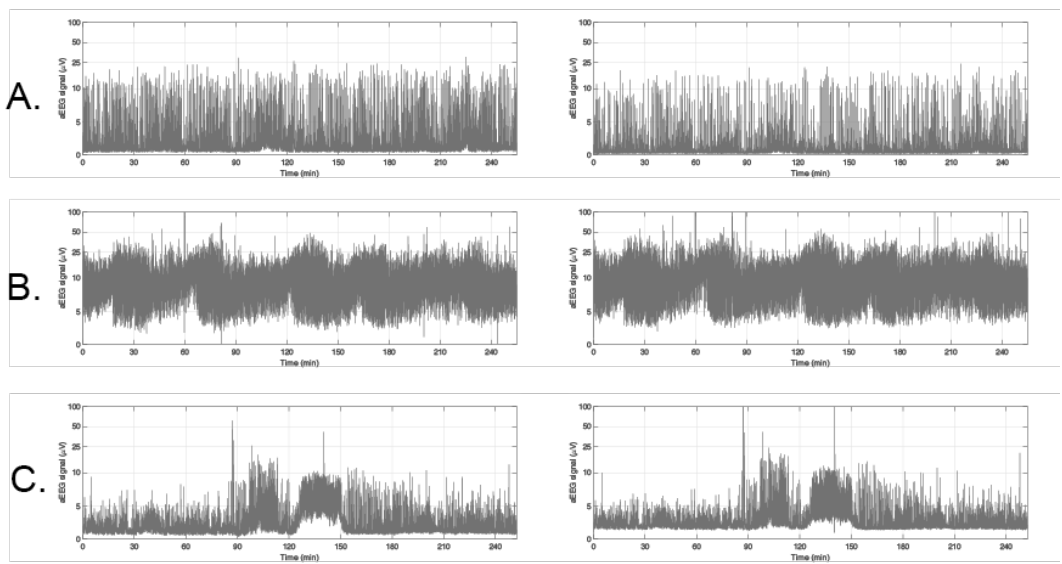

Fig. 4. Example of paired aEEG signals calculated by the AnalyZe reference algorithm (left column) and the WU-NEAT algorithm (right column). Row A is a burst-suppression pattern from a 2-day old 24-week EGA infant. Row B is a continuous pattern with evidence of cyclicity from a 4-day term gestation infant, previously suffering from hypoxic-ischemic encephalopathy, now resolved. Row C is a 2-day old term gestation infant suffering from hypoxic ischemic encephalopathy with a predominant burst-suppression pattern with a 30-minute seizure noted 120 minutes into the recording

A closer match may be found in "Neural," a package of MATLAB scripts released by O'Toole and colleagues [19]. Provided are tools to calculate the spectral edge frequency, the range EEG (similar to aEEG), and the interburst interval. While certainly of interest to many researchers, this toolset lacks a GUI, is missing a method for calculating the aEEG, does not have a standardized file format, and has not been validated against an FDA-approved standard.

The analytic tools provided in WU-NEAT compromise the most commonly used methods for analysis of limited-channel EEG in current clinical practice and research. These tools are encapsulated in an easy to use GUI, which is accessible to investigators of different technical aptitudes. The included components, aEEG and SEF, are the most utilized techniques for limited-channel neonate EEG analysis.

WU-NEAT represents an important first step in re-establishing this neglected field. This extensible framework is now available for the next stage—the creation of novel, innovative tools. The modular and flexible nature of WU-NEAT makes the addition of other algorithms a trivial process, particularly when considering the near-impossibility of such a goal without first developing this foundation.

Limited-channel EEG has been long been plagued questions of inter-rater reliability, in no small part due to differences in hardware and software display. While this issue is less problematic for qualitative scoring systems (such as the Hellström-Westas or Burdjalov systems), it is a critical problem for quantitative analysis. Through the clinical validation provided in this study, users can be confident that the use of WU-NEAT will provide comparable output to the current, FDA-approved commercial standard.

WU-NEAT represents a key step in the establishment of a widely-available, standardized platform for quantitative research using limited-channel EEG recordings from neonates. It is our hope that WU-NEAT will be used to develop innovative metrics which can inform or even predict neurocritical events with the aim of ameliorating or even preventing neurologic injury.

## V. Conclusion

An open source implementation of the most common limited-channel EEG analysis techniques (aEEG and SEF) is possible. The resulting algorithms generate numerically equivalent output to the FDA-approved reference standard. Testing with blinded reviewers shows that the output is essentially indistinguishable in clinical use.

Acknowledgment

The authors wish to acknowledge the efforts of Anthony Barton and Laura Atwood for their exceptional efforts at patient recruitment, monitoring, and data collection.